\documentclass[conference]{IEEEtran}
\IEEEoverridecommandlockouts

\usepackage{cite}
\usepackage{amsmath,amssymb,amsfonts}
\usepackage{algorithmic}
\usepackage{graphicx}
\usepackage{textcomp}
\usepackage{xcolor}
\def\BibTeX{{\rm B\kern-.05em{\sc i\kern-.025em b}\kern-.08em
    T\kern-.1667em\lower.7ex\hbox{E}\kern-.125emX}}
\begin{document}

\title{Bug Searching in Smart Contract\\
}

\author{\IEEEauthorblockN{1\textsuperscript{st} Xiaotao Feng}
\IEEEauthorblockA{\textit{Swinburne University of Technology} \\
Melbourne, Australia \\
101973718@student.swin.edu.com}
\and
\IEEEauthorblockN{2\textsuperscript{nd} Qin Wang}
\IEEEauthorblockA{\textit{Swinburne University of Technology} \\
Melbourne, Australia \\
qinwang@swin.edu.au}
\and
\IEEEauthorblockN{3\textsuperscript{rd} Xiaogang Zhu}
\IEEEauthorblockA{\textit{Swinburne University of Technology} \\
Melbourne, Australia \\
xiaogangzhu@swin.edu.au}
\and
\IEEEauthorblockN{4\textsuperscript{th} Sheng Wen}
\IEEEauthorblockA{\textit{Swinburne University of Technology} \\
Melbourne, Australia \\
swen@swin.edu.au}
}

\maketitle

\begin{abstract}
With the frantic development of smart contracts on the Ethereum platform, its market value has also climbed. In 2016, people were shocked by the loss of nearly \$50 million in cryptocurrencies from the DAO reentrancy attack. Due to the tremendous amount of money flowing in smart contracts, its security has attracted much attention of researchers. In this paper, we investigated several common smart contract vulnerabilities and analyzed their possible scenarios and how they may be exploited. Furthermore, we survey the smart contract vulnerability detection tools for the Ethereum platform in recent years. We found that these tools have similar prototypes in software vulnerability detection technology. Moreover, for the features of public distribution systems such as Ethereum, we present the new challenges that these software vulnerability detection technologies face.
\end{abstract}


\section{Introduction}

Decentralized cryptocurrencies have gained tremendous attention from both academia and industry. The emerging novel technology originated from these systems is blockchain, a sequentially ordered ledger system. The system possesses the properties of being distributive, irreversible, unforgeable, and traceable. Ethereum \cite{EthWP19}, as the most accessible blockchain platform, supports distributed applications in different scenarios through the underlying online virtual machine called EVM, which is a fundamental layer for the complete execution of the smart contracts. Smart contract \cite{EthSC19}\cite{Atzei2017} is a collection of code and data (also known as states) executing on the blockchain system. It is Turing-complete which allows us to write the pre-defined rules. Smart contract is pretty suitable for the scenarios requiring dependable security, irreversible persistence, and high trusts, such as the digital assets, online voting, gambling games, insurance, property managements, and financial applications.

However, there are many vulnerabilities in smart contracts \cite{Ring}\cite{Tsankov:2018:SPS:3243734.3243780}\cite{Nikolic:2018:FGP:3274694.3274743}\cite{tsankov2018securify}\cite{liu2018reguard}\cite{8429306}\cite{Town}, and the high financial status brings higher interaction risks. Unlike traditional distributed application platforms, smart contract platforms such as Ethereum allow anyone to join. This high openness makes the EVM environment very vulnerable. Also, some vulnerabilities can only be exploited in some of Ethereum's unique new features (such as timestamp design, gas settings and fallback function). For example, The DAO\cite{DAO} exploits a variety of well-documented reentry attacks, resulting in the theft of Ethernet worth more than \$50 million. As a result, the security issue of smart contracts on Ethereum has attracted much attention.

In the world of vulnerability detection, researchers have developed many tools to find vulnerabilities in programs. We focus on three tools, including fuzz testing \cite{Miller1990empirical}, formal verification, and symbolic execution, that have already found many vulnerabilities. However, as for such tools applying in the smart contract, it is still at the beginning of the research. The nature of fuzzing is to generate inputs for programs and tries to find vulnerabilities based on the results of executing programs \cite{Miller1990empirical}. The biggest challenge of fuzzing comes from the fact that it tests a program in random behavior. Formal verification is a method based on logical deduction and tries to prove or disprove the correctness of a program. Researchers have to prove their conclusions strictly so that the target program can be fully trusted or abandoned. The last tool we introduce is symbolic execution, which treats variables in a program as symbolic values. It regards each condition in the program as a constraint and tries to find a possible solution along with an execution path. The main challenge of symbolic execution is path explosion which results from loops or arrays.  When such tools are applied in smart contract, researchers have to figure out specific methods to fit tools into smart contracts. The details will be discussed in the following.

In this paper, we survey some common vulnerabilities on EVM and their triggering mechanisms. We also introduce some of the tools in the software vulnerability detection industries on this new platform and their work-flow as well as features. Further, for some new features of the EVM platform, we also present some new challenges for these software vulnerability detection tools.

\section{Overview of EVM}

\subsection{EVM Model}

Ethereum Virtual Machine (EVM) provides a practical environment for the execution of smart contracts in the distributed Ethereum platform. It refers to a complete suite of logic processes of deploying, compiling, and executing \cite{EthWP19}. Users can make their contracts automatically be executed according to the pre-defined rule via transaction-based state transitions \cite{EthSC19}. To achieve the state machine transition, the smart contract can be seen as one type of shared-state. The globally shared-state is made up by small units, called $account$, which can interact with each other through messaging. Each account associates with a state and a 20-byte address as the identifier. There are two different types of accounts:
\begin{itemize}
    \item \textit{Externally owned account}, controlled by a private key without any associated code.
    \item \textit{Contract account}, controlled by the corresponding contract code, which sets the actions and operations.
\end{itemize}

\begin{figure}[!t] 
\includegraphics[width=1.0\columnwidth]{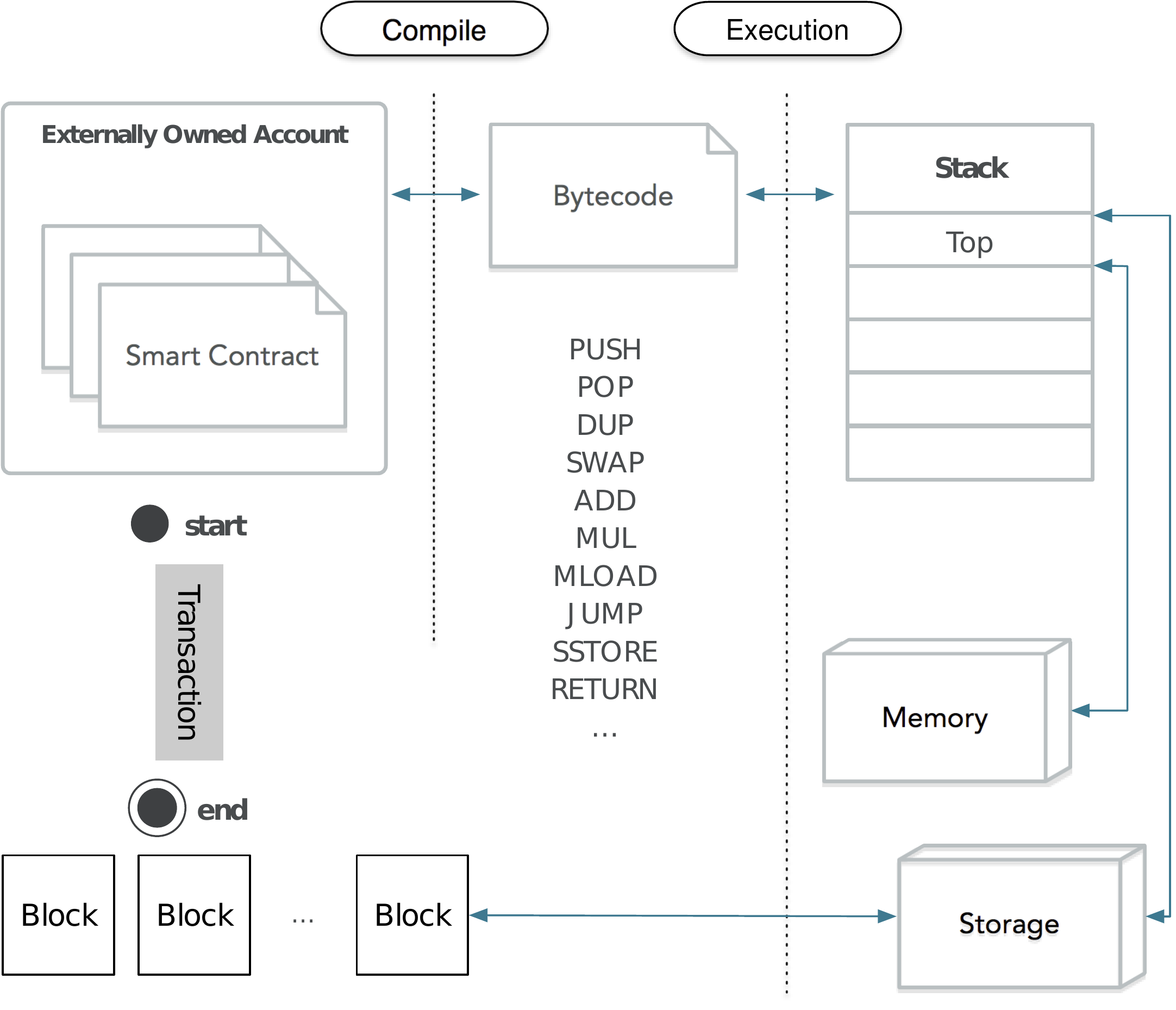} 
\caption{EVM Model}
\label{EVM}
\end{figure}

The externally owned account sends the message to another externally owned account or contract account by creating and signing the transaction under the private key. The message transmitted between two externally owned account is just a simple value transfer. But a message from an externally owned account to a contract account will activate the code in the contract account, performing the corresponding actions (such as transferring tokens, logging storage, generating new contracts, minting new tokens, calculating values, etc.). Unlike an externally owned account, the contract account cannot initiate a transaction by its own. Instead, the contract account triggers the transaction only after receiving a transaction. The EVM model is presented in the Fig.\ref{EVM}.

There are four components in both types of the accounts:
\begin{itemize}
\item \textit{Nonce:} It represents the transaction number sent by the externally owned account, and it also represents the contract number created by the contract account.
\item \textit{Balance:} It means the remaining value of the address, dominated as $Wei$ where 1 Ether=$10^18$ Wei.
\item \textit{Root:} The root represents the hash value of the Merkle Patricia tree which encodes the Superimposed hash of the stored content in each account. 
\item \textit{CodeHash:} It is specially used for contract account, where the code of smart contract is saved as codeHash. For externally owned accounts, this field is an empty string.
\end{itemize}

\subsection{Features in EVM}

\paragraph{Gas As Fee}
Every defined calculation generated by transactions requires an amount of cost to be paid. Gas is used to measure the unit of cost in each specific calculation. Correspondingly, Gas price is the value spent on each gas, measured by ``gwei'' where 1 gwei=1,000,000,000 wei. For each transaction, the sender needs to set two parameters including gas limit and gas price. The total amount represents the cap that the sender is willing to pay for the execution of the transaction. The total amount is calculated by the gas limit and the gas price limit as follow. 
\[ \textrm{Gas Limit} \times  \textrm{Gas Price} = \textrm{Total amount} \]

Any unused gas at the end of the transaction will be returned to the sender for redemption. If there does not exist enough gas to execute the transaction, the transaction will be considered invalid. In this case, the process terminates where all changed states return back to the initial states. Since the system is still working on calculation before running out of gas, no gas will theoretically be returned back to the sender. Instead, all fees used in execution are sent to the miners as the reward who have already made an effort to calculate and verify the transaction.

\paragraph{Fallback Function in Solidity}

The \texttt{fallback} function, is an uniquely unnamed function in the solidity. It is automatically executed only when no other function matches the specified function identifier \cite{EthSC19}. At the same time whenever the address of the contract receives plain Ether without messages, the function can be called. \texttt{fallback} function has no arguments or return nothing. More specifically, when we transfer tokens without any readable data via the function \texttt{address.send(ether)}, the contract will automatically execute the \texttt{fallback} function to make the transition of state proceed. To make the \texttt{send} outputs TRUE, the fallback function must be marked $payable$.

Since the \texttt{send} function always calls fallback, it is dangerous to be attacked by malicious attackers such as DAO \cite{DAO}. Especially in the scenario of dividend where the \texttt{send} operation is deployed on a series of accounts，if there exists at least one malicious account holding \texttt{fallback} functions to infinite loop, it will cause all \texttt{send} processes to fail where the gas is used out. In order to solve this problem, the \texttt{send} function sets a limited 2300 gas as the maximum even if the gas is sufficient. Therefore, except for the operations such as log in the fallback function, one can hardly do anything to keep the liveness of the whole system.

\paragraph{Timestamp}
Timestamp, as one kind of identifier of blocks and transactions, contained in each block header in the form of Unix time. In addition, the irreversible timestamp avoids the faking of blocks by adversaries. Based on timestamp, blockchain system confirms that each block is sequentially connected. The timestamp proves the sequence of events in which no one can tamper with it. Timestamp can be seen as the role of notary in blockchain, which is credible for the public participants.

\section{Vulnerabilities}

In this section, we will discuss three common vulnerabilities in smart contract and present the examples of attacks. 


\subsection{Reentrancy}

Recursion is a widespread logical processing method in traditional programming languages, but this operation is likely to become a vulnerability in Solidity. 

\begin{figure*}[!t]
\centering
\includegraphics[width=1.0\textwidth]{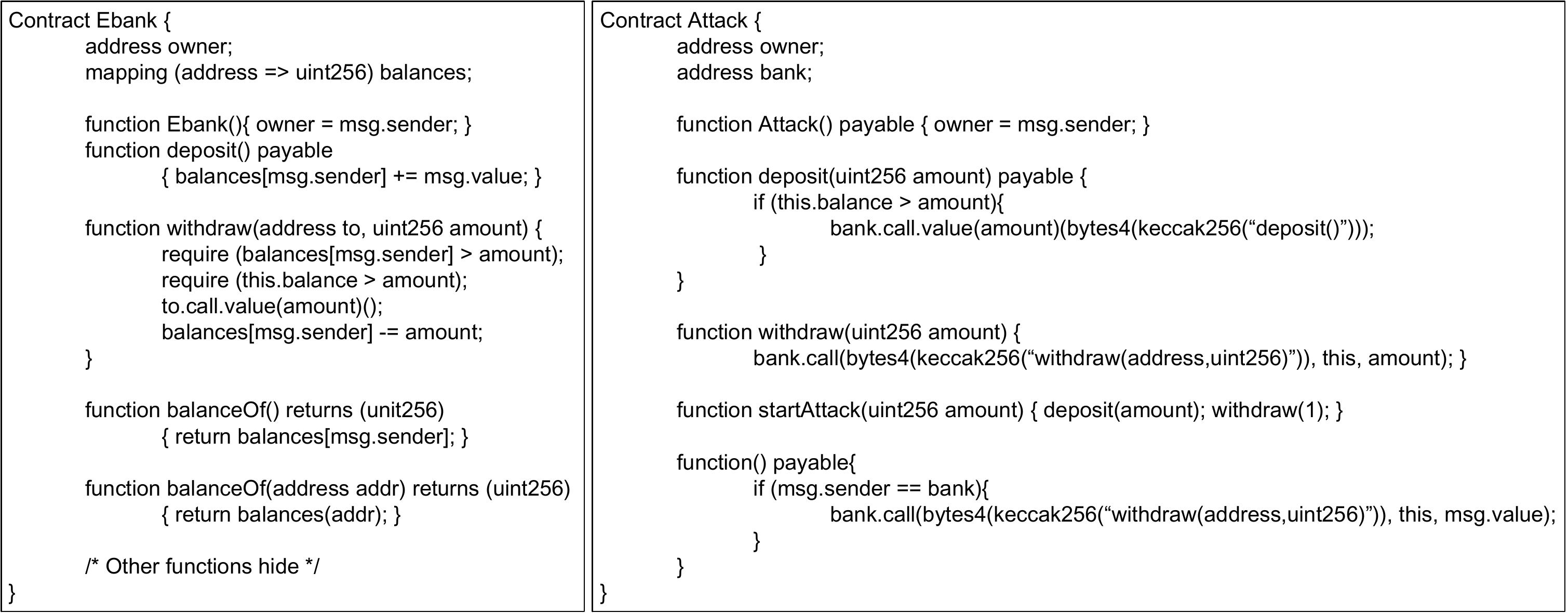}
\caption{Ebank and Attack contracts}\label{fig-ENA} 
\end{figure*}

Fig. \ref{fig-ENA} Ebank contract shows the implementation of a contract for a public bank. Any user can deposit Ether to \textbf{EBank} and the contract records the Ether of each account. In this scenario, when the contract withdraws the saving (\textbf{withdraw(address to, uint256 amount))}, it ensures that the account has enough balance (\textbf{require (balances[msg.sender] \textgreater  amount)}) and that the bank has sufficient funds for the withdrawal. If the above two conditions are satisfied, the contract will send the Ether to user(\textbf{to.call.value(amount)()}) and change the balance of corresponding value(\textbf{balances[msg.sender] -= amount}).


In programming languages such as C and C++, the code in Fig.\ref{fig-ENA} Ebank contract can be run correctly. However, such piece of code may be vulnerable in Ethereum`s smart contract due to its own grammar. In Fig.\ref{fig-ENA} Ebank contract, the contract utilises \textbf{call.value()()} to send user Ether. Distinguished from the functions \textbf{send()} and \textbf{transfer()}, function \textbf{call.value()()} gives all the rest of gas to external call (fallback function). If the target address is a contract address when making an Ether transaction, the contract`s fallback function is called by default.

To show how attackers can exploit this vulnerability, we design an attack contract in \textbf{Fig.\ref{fig-ENA}}. In \textbf{startAttack()}, the contract firstly \textbf{deposit()} the specified amount tokens in bank, and one token is taken through the \textbf{withdraw()}. Because the bank contract uses \textbf{call} function to send Ether to the target, this will call the target contract`s \textbf{fallback} function which calls the \textbf{withdraw()}. The user‘s balance is modified after the Ether is transferred, the balance on the attacker’s account remains unchanged and the attacker can always take out Ether from the bank. Therefore, the two contracts of bank and attack fall into a recursive state until that Ether in the bank is sent to the attacker or attacker stops the contract.

\subsection{Gasless}

The EVM usually sets the gas limit at 2,300, and the miners who are good at calculating can use the delicate contract structure combined with fallback function to run out of gas. This will result in an error in the running contract so that the miner can get profits from it or achieve other goals.

\begin{figure}[!t] 
\centering
\includegraphics[width=0.85\columnwidth]{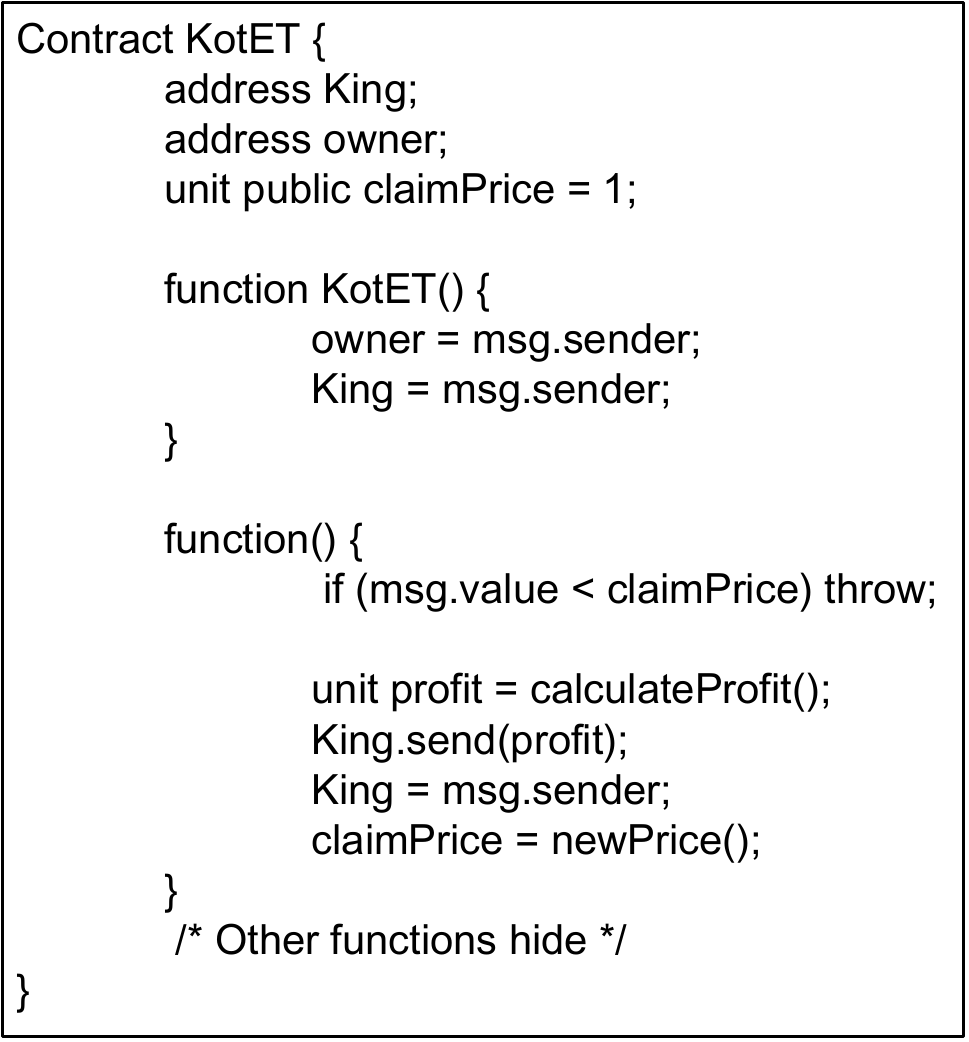} 
\caption{KotET contract}\label{kotet}
\end{figure}

There is a game called \textbf{KingOfTheEtherThrone} \cite{KotET} and the game is played by sending Ether to a smart contract called KotEt(as shown in Fig.\ref{kotet}). Players who want to be king must pay some Ether to the current king, plus a small amount of fee to KotET contract. Then, the king will get profit from the difference between the price he paid for the throne and the price other player pays to be a new king. 

Supposing a player wants to be king, he wants KotET to send a certain amount(\textbf{msg.value}) of Ether. The fallback function of KotET is called and it will check if \textbf{msg.value} is greater than the quote for the previous king setting. If it is less than the quote for the previous king setting (i.e., the auction failed), it will be abandoned. On the contrary, the player will get the throne and become the new king. 

This contract seems to be fine, but there will be a gasless \texttt{send} bug. When \textbf{king.send(profit)} fails to execute (gas is not enough to execute fallback()), the throne will be held by this contract.

\subsection{Timestamp}

In Solidity language, it defines many block state variables\cite{Global} like timestamp, random seed and block number. Since these state variables are written at the head of each block, the malicious miner may modify it and get profit from it. These block state variables can make the Ether flows along different program paths. Here we use timestamp to illustrate how such a vulnerability is exploited by malicious miners.

Block timestamps have traditionally been used for a variety of applications, such as functions for random numbers, locking contract for a period of time, and various conditional statements based on time-varying states. Miners have the ability to adjust the timestamp slightly, and if the block timestamp is misused in a smart contract, it can prove to be quite dangerous.

\begin{figure}[!t] 
\centering
\includegraphics[width=0.85\columnwidth]{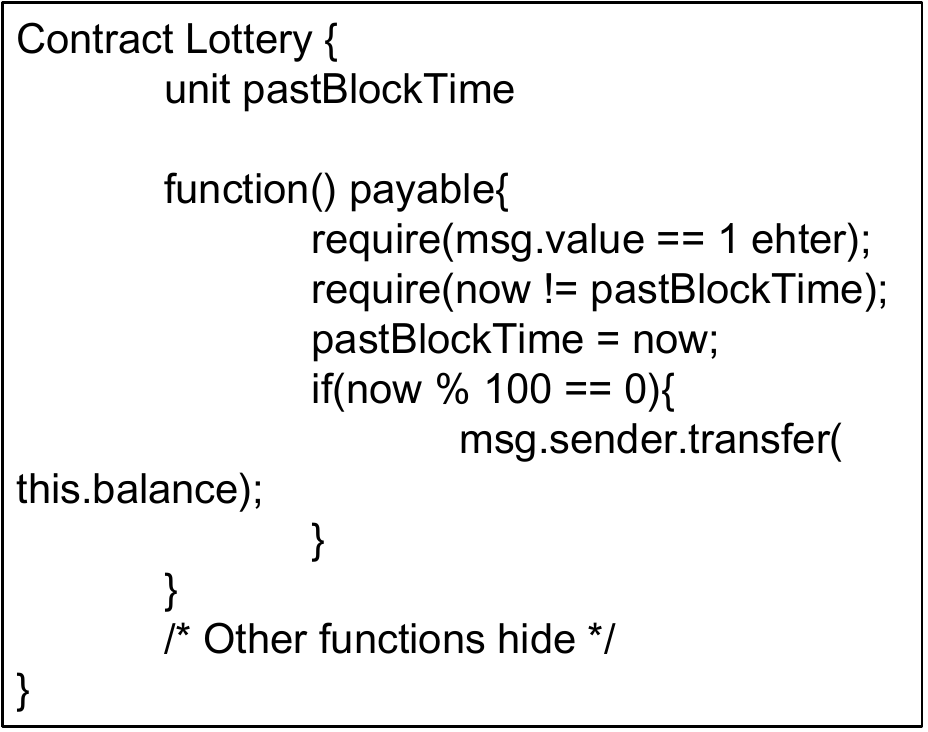}
\caption{Lottery contract}\label{fig-Lottery}
\end{figure}

\textbf{Block.timestamp} (or \textbf{now}) can be manipulated by miners if they have incentives to do so. We build a simple contract that is vulnerable to exploitation by miners (Fig.\ref{fig-Lottery}).

In the Fig.\ref{fig-Lottery}, this contract is a simple lottery. Each block has a trade to bet 1 Ether and get the chance to win the entire balance in the contract. The assumption here is that the last two digits of block.timestamp are evenly distributed. If so, there will be a 1\% chance to win this lottery.

However, as far as we know, miners can adjust the time stamp according to their wishes. In this particular case, if there is enough Ether in the contract, the miner who digs out a block will be motivated to choose a \textbf{block.timestamp} (or \textbf{now}) to 100 with a timestamp of 0. In doing so, they may win Ether and block rewards in this contract.

\section{Tools}

\subsection{Fuzzing}

\begin{figure}[!t] 
\includegraphics[width=1.0\columnwidth]{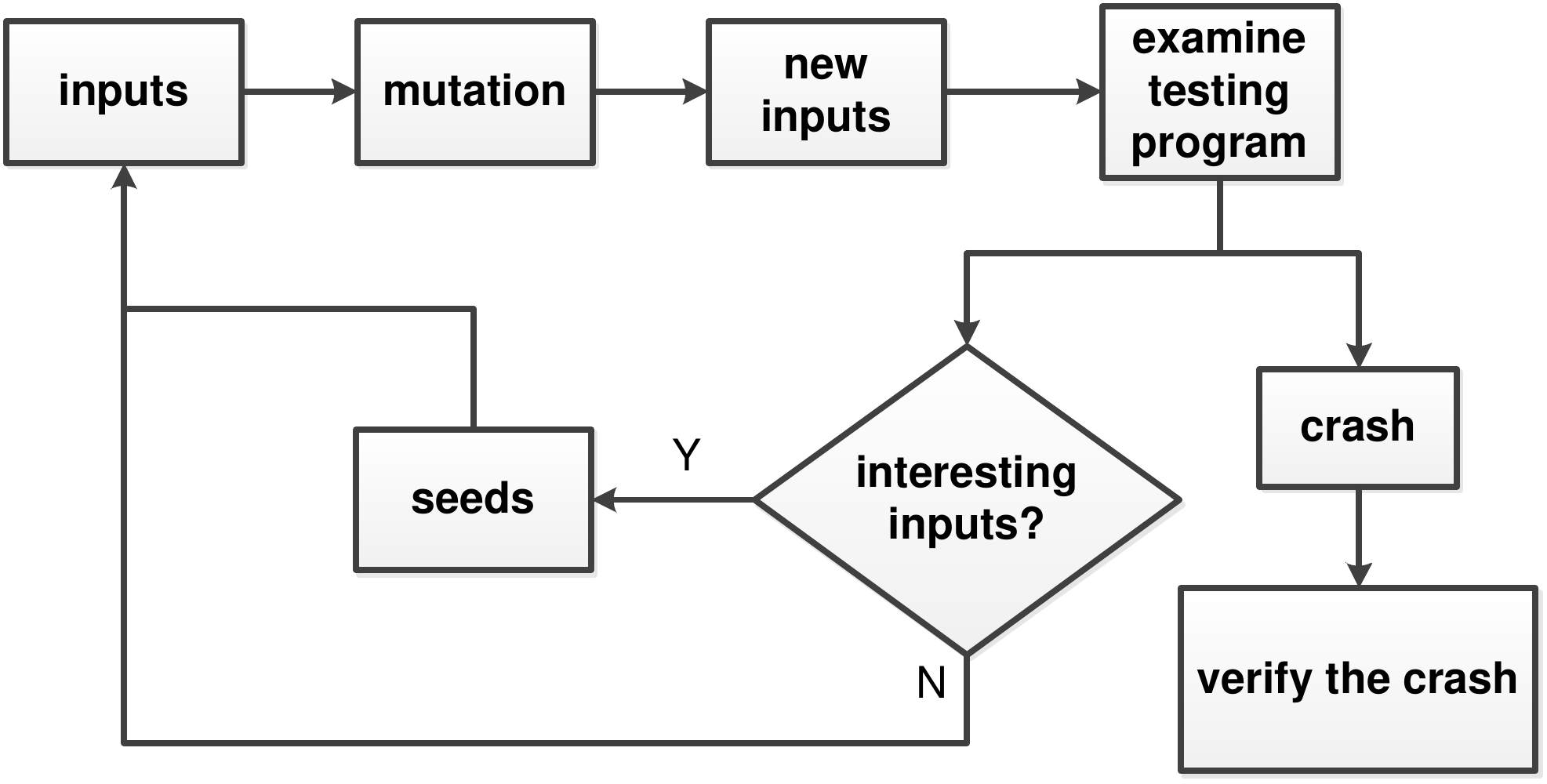} 
\caption{The typical procedure of fuzzing. Original inputs are mutated into new inputs, and these new inputs will examine the testing program. The aim is to crash the testing program, and then check whether the crash is a bug.}\label{fig-fuzzing-proc}
\end{figure}

Fuzzing \cite{Miller1990empirical} is a technique that randomly generates inputs to examine testing programs. In most situations, fuzzing intends to crash a testing program. Moreover, when it crashes, we can check whether the crash is a bug or not.
Fig.\ref{fig-fuzzing-proc} shows the typical procedure of fuzzing.
At the very first execution, fuzzing needs original inputs to be mutated. Then, new inputs are mutated and generated from the original inputs, which will examine the testing program. After the examination, fuzzing will check whether the new inputs are interesting or not. The interesting inputs are saved to be seeds, which will be chosen as inputs. If the testing program crashes, we have to verify whether the crash is a bug.

\begin{figure}[!t] 
\includegraphics[width=1.0\columnwidth]{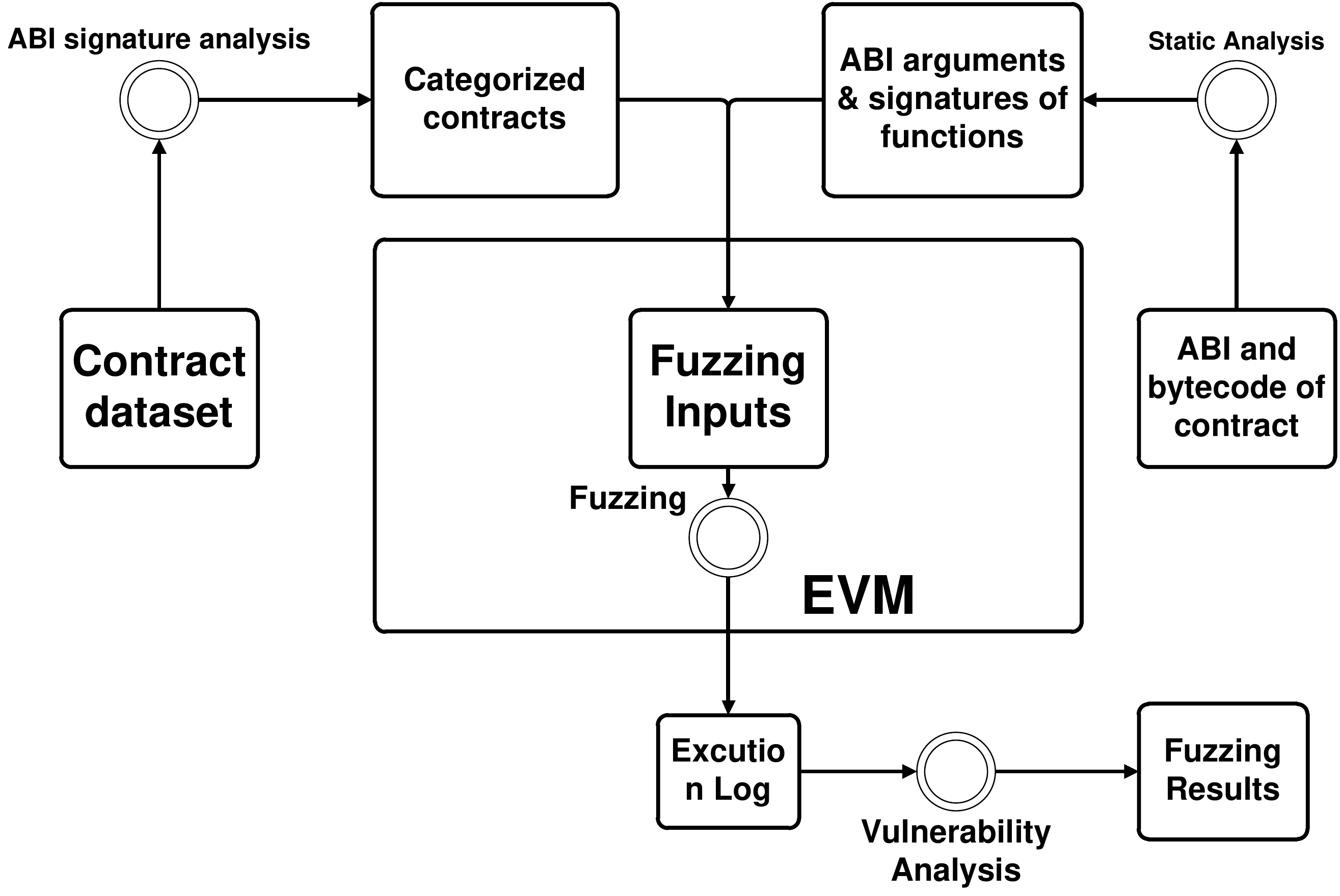} 
\caption{Overview of the ContractFUzzer Tool}\label{fig-contra-fuzz}
\end{figure}

In Jiang's work\cite{jiang2018contractfuzzer}, they create a new fuzzer, named ContractFuzzer, which is a novel fuzzer to fuzz Ehereum smart contracts. ContractFuzzer is vulnerability detecting tool which is built based on traditional fuzzing combining with static analysis. Fig.\ref{fig-contra-fuzz} shows how ContractFuzzer works on fuzzing smart contract. It provides an offline \textbf{EVM instrumentation tool} which can monitor the execution of smart contracts for subsequent analysis. The ContractFuzzer firstly works on analyzing the \textbf{ABI interface and bytecode} of the smart contract which is collected in \textbf{contract dataset}. Then the ABI function arguments and signatures will be extracted for the \textbf{ABI signature analysis}. ContractFuzzer generates the input seed for online fuzzing based on the two types of analysis, and finally, it starts to do fuzzing test and detect security vulnerabilities via analyzing \textbf{execution logs}.

\subsection{Formal Verification}

In the context of hardware and software systems, formal verification is the act of proving or disproving the correctness of intended algorithms underlying a system concerning a certain formal specification or property, using formal methods of mathematics. 

Compare with the dynamic detection methods like fuzzing, for Ethereum, static methods such as the formal verification does not require a simulated execution environment and provides better precision and false positive rate in vulnerability analysis \cite{OYENTE}. In general, automatic formal verification can be divided into three main types: 1) Automated theorem proving, 2) Model checking and 3) Abstract interpretation. For smart contract platforms, model checking is appropriate because of the smaller size of smart contracts.

\begin{figure}[!t] 
\includegraphics[width=1.0\columnwidth]{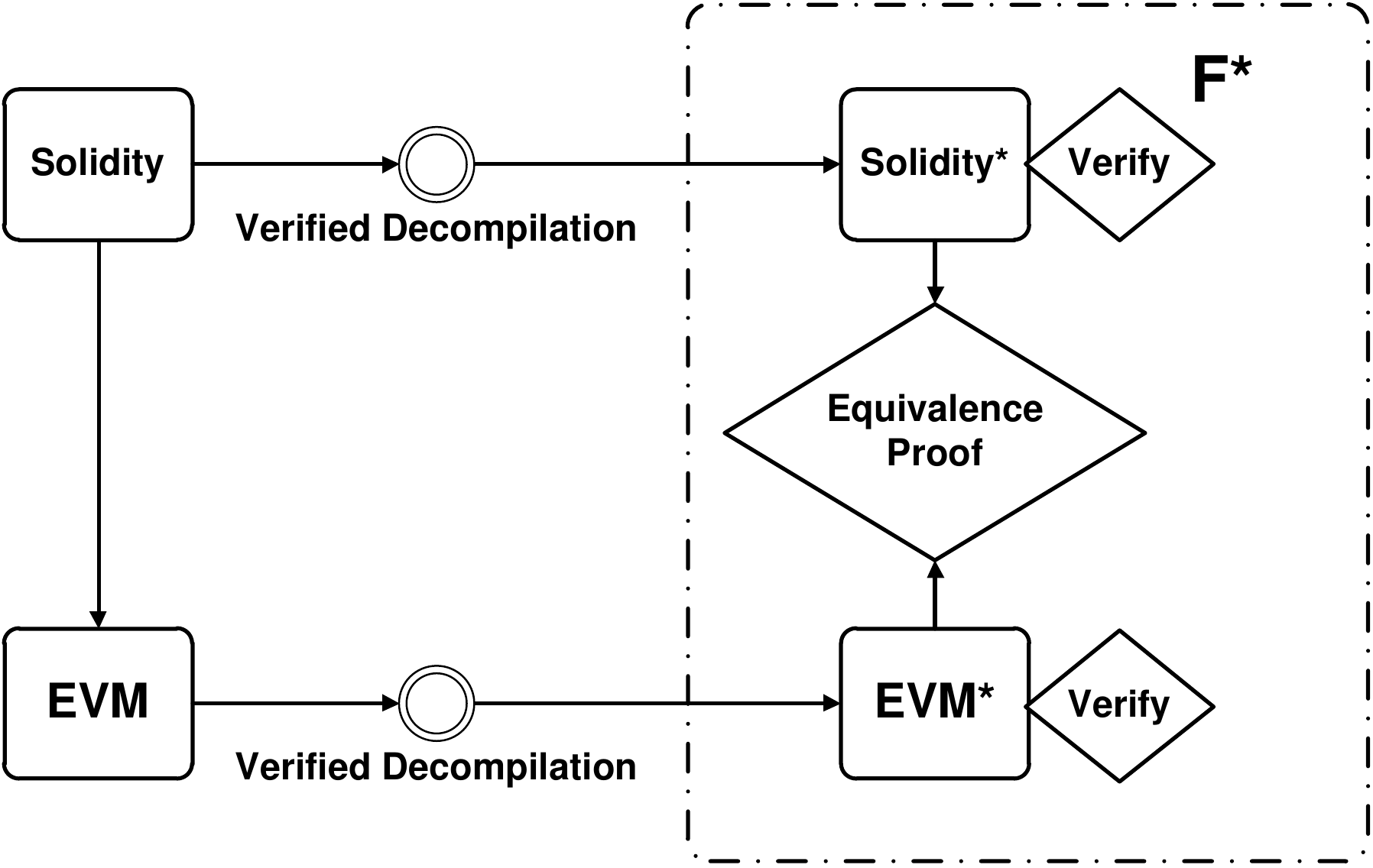}
\caption{Outline of our verification architecture}\label{fig-verif-arch}
\end{figure}

Inspired by the process of processing JavaScript\cite{swamy2014gradual}, Bhargavan`s team\cite{bhargavan2016formal} outlines a framework to verify Ethereum smart contracts using formal verification. After translating and decompiling the Solidity code and EVM bytecode into a functional programming language named F*, they will determine the existence of a vulnerability in the contract by verifying the equivalence of the two in the F* language results (See Fig.\ref{fig-verif-arch} for details).

However, there are still significant limitations to this approach. Bhargavan also mentioned in their evaluation part that this language-based process cannot support many Solidity language features. It can only translate and typecheck 46 out of the 396 contracts they collected from https://etherscan.io.



\begin{figure*}[!t]
\centering
\includegraphics[width=0.9\textwidth]{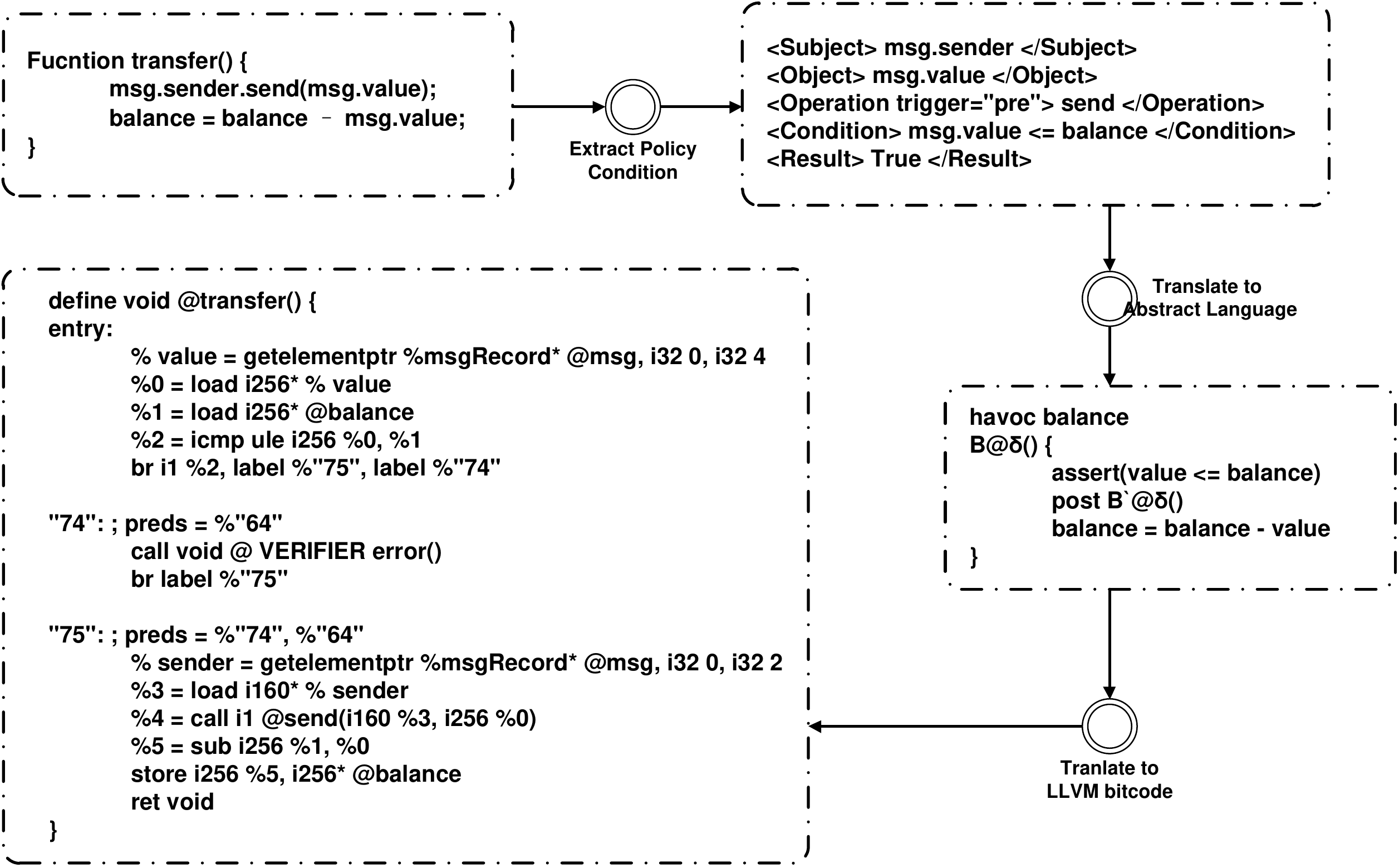}
\caption{An example of Zeus working flow}\label{fig-zeus-wf}
\end{figure*}

ZEUS\cite{kalra2018zeus} did more work on translating the smart contract language. Zeus consists of three parts: a) policy builder, b) source code translator, and c) verifier. The policy builder performs static analysis on the smart contract code, extract the predicate from the policy condition and then insert the assertion into the contract code. Unlike \cite{bhargavan2016formal}, Zeus does not directly deal with the source code of Solidity but converts it into LLVM bytecode by the source code translator. Finally, the verifier will check the assertion inserted by the policy builder before, and then determine violations. Fig.\ref{fig-zeus-wf} is an example to explain how the Zeus works on smart contract. Firstly, ZEUS formalises Solidity Semantics into Abstract language. For the condition policy in these codes, ZEUS creates an XACML-Styled \cite{XACML} five-tuple policy specification to describe and convert these policies into assert statements. The LLVM translator then helps ZEUS convert these solidity codes into LLVM's bytecode and finally validate the code with the CHC\cite{McMillan:2007:ISM:1763048.1763057} symbolic model checker.


\subsection{Symbolic Execution}
Symbolic execution is a technique based on formal verification. It analyses programs to test whether specific properties can be violated. This technique can yield strong guarantees on the checked property due to the nature of symbolic execution, which is it can simultaneously explore multiple paths. The key idea of symbolic execution is to analyze programs based on symbolic values, rather than concrete input values.

In symbolic execution, the execution part\cite{baldoni2018survey}, which is performed by an engine, maintains each explored control flow path. This engine contains two parts: 1) a first-order Boolean formula describes the conditions included in branches of paths, and 2) a symbolic memory store maps variables to symbolic expressions. Branch execution updates the formula while a model checker verifies whether there are any violations.

\begin{figure}[!t] 
\includegraphics[width=1.0\columnwidth]{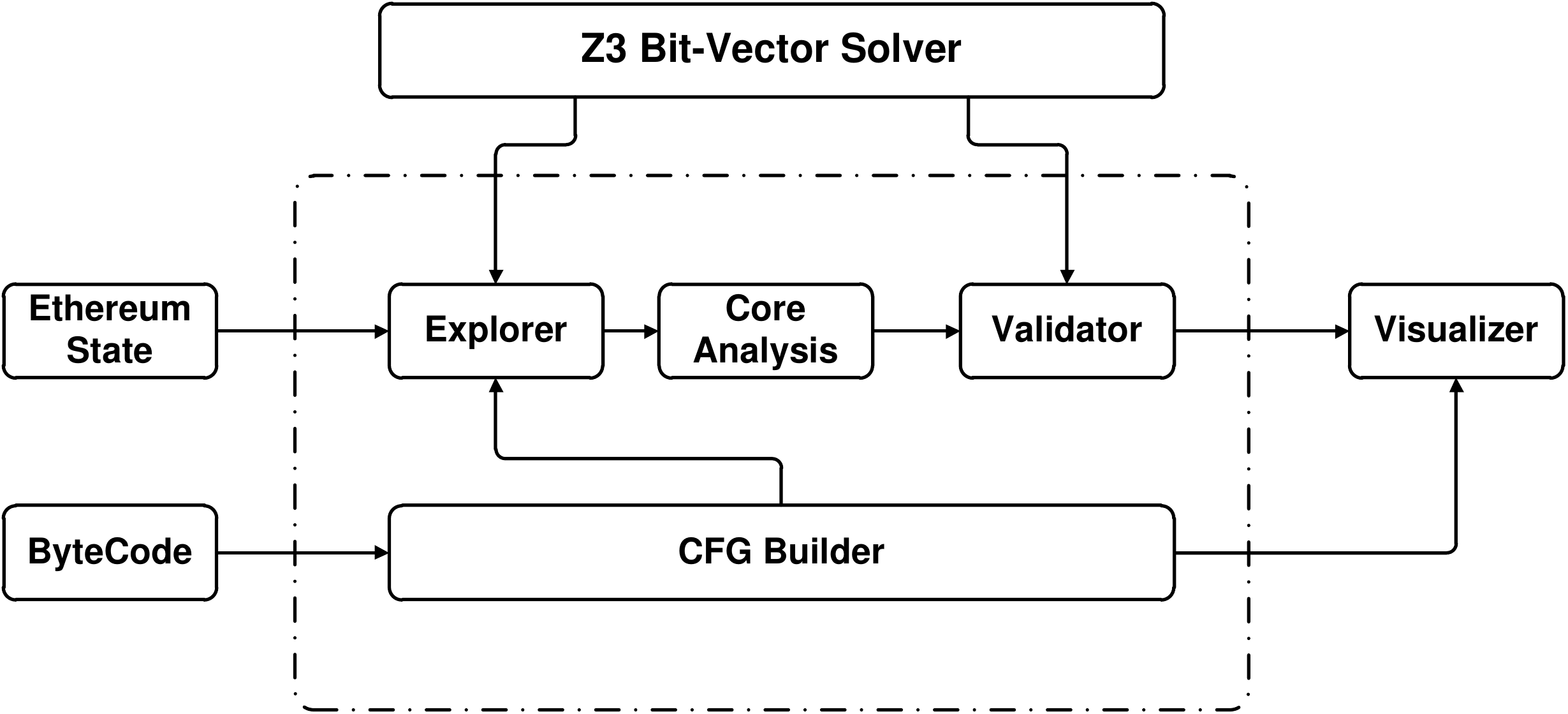}
\caption{Overview of the OYENTE Tool. Oyente was built based on modular design, and it consists of four main components: 1) \textbf{CFG Builder} draws a control flow graph of whole \textbf{Byte Code}; 2) \textbf{Explorer} is an interpreter loop, and it runs a single instruction when it gets a state from last the run; 3) \textbf{Core Analysis} contains several sub-components that help it analyze different security vulnerabilities; 4) \textbf{Validation} helps OYENTE remove false positive via manually verifying the results provided by \textbf{Core Analysis}.}\label{fig-oyente} 
\end{figure}

Luu \cite{OYENTE} created a static analysis tool named OYENTE based on symbolic execution technique to help developers avoid the vulnerabilities in writing smart contracts. Fig.\ref{fig-oyente} is an overview of OYENTE. After extracting the smart contract bytecode from the EVM, \textbf{CFGBuilder} plots the main control flow graph (CFG) for each smart contract. Then, \textbf{Explorer} performs the Symbolic execution on these CFGs, and uses the Z3 solver \cite{Z3} to complete the CFG block entry condition according to the path constraints. \textbf{Core Analysis} performs a vulnerability analysis on the collated CFGs, and at the end \textbf{Validation} will verify the analysis results.


\subsection{Language Translation}
Language translation is a common support tool for software vulnerability detection. In general, it can transform some hard-to-read or unpopular languages into some well-known programming languages. For the static method of vulnerability analysis, if you need to analyze an executable file, the general practice is to convert it into a high-level language that we know well with the decompilation tool \cite{ida} for the researchers to further processing. This kind of decompilation is a translation of assembly language into high-level languages such as C, C++, etc. Similarly, disassembly can help researchers solve problems on binary files, and in essence, it is also a translation technique. This translation technology can be used not only in different levels of programming language but also help us achieve the conversion of peer languages, such as C++ to JAVA\cite{language}. This kind of translation is not meaningless. It can turn some hard-to-read languages (e.g., machine code) into popular languages with many supporters, and then the program will be easier to understand, and researchers can use some sophisticated analytical techniques according to the language. 

As a new language with the birth of new technologies, the programming language environment for smart contracts is far less sophisticated than other mature languages like C++ and so on. As a result, many research teams have chosen to translate the language of smart contracts into the form they want, and then use existing sophisticated software analysis techniques to detect the vulnerability problems in smart contracts. Bhargavan`s team\cite{bhargavan2016formal} translates Ethereum and smart contracts into F*, which is a functional programming language designed for program verification, to analyze the security and functional correctness of the platform while it is running. Similarly, in ZEUS's\cite{kalra2018zeus} work, they implemented a tool to translate solidity code into LLVM bytecode and proposed a model detection scheme based on LLVM bytecode. Leveraging LLVM bytecode helps their analytical work take advantage of the support of robust industry tools on LLVM platform. Brent et al. \cite{brent2018vandal} used the decompilation method. They decompiled and analyzed the bytecode of the smart contract extracted from Ethereum, and got readable low-level mnemonics that are annotated with program counter addresses and help Vandal\cite{brent2018vandal} generate the control flow graph of these smart contracts.

\section{Challenge}

As we mentioned in this paper, software vulnerability detection methods have been well applied to smart contract platforms. However, due to the many differences among the smart contracts, the traditional software programs, and the bottlenecks of these vulnerability detection technologies, we still face challenges in the detection of vulnerability in smart contracts.


\subsection{Fuzzing}



Due to the concurrency of smart contracts and the distributed design of blockchain, fuzzing on smart contract platforms has many differences from fuzzing in the traditional sense. In software vulnerability mining, whenever fuzzing causes the target program to crash once, we consider it a potential vulnerability (requires tools or manual verification). In smart contracts, the vulnerability we mentioned earlier is almost impossible to crash the EVM, even though the collapse of a single contract does not affect the entire EVM. Therefore, as we mentioned in ContractFuzzer, when using fuzzing in smart contracts, we also need to record the results of each execution result of fuzzing and perform additional analysis to verify vulnerabilities. ContractFuzzer uses a predefined test oracle to solve this problem.

For example, for a gas-free problem, ContractFuzzer will repeatedly execute a single contract and check if the residual gas value of the \textbf {send()} function in the result analysis is zero. However, the results of some single contract executions do not reveal the vulnerability which requires special circumstances (such as reentrancy attacks, which require interactive calls between two smart contracts). In \cite{jiang2018contractfuzzer}, for the reentrancy attack, they created three kinds of accounts and two options for \textbf{call.value()}. In this scenario, if a single contract is to be fuzzed \textit{k} times, the ContractFuzzer will perform 6(2*3)*\textit{k} times fuzzing, and then analyze the results. For this kind of reentry attack, we can make this scene design more complicated to cover more situations, but at the same time, the fuzzing execution time and analysis time will be accompanied by complexity growth. Therefore, how to find the balance between scene complexity and vulnerability patterns coverage is a challenge for fuzzing method.  


Similarly, some of the problems that appear in the software fuzzing test also exist in smart contracts. Sanity checks (like magic number or checksum condition checks) in programs has always been a challenge for fuzzing, and its presence has also increased the false positive rate of fuzzing experiments. Due to the input generated by fuzzing has a high degree of randomness, fuzzing is difficult to pass for a condition check such as `Str == ``HelloWorld''', resulting in some paths becoming difficult or inaccessible. This problem has become more severe in smart contracts. Since many smart contracts include external attributes as part of the verification (such as timestamps), this leads to the fact that if the environment at this time does not meet the requirements of the contract (the date limit has passed). Then, the contract may not be able to be entered, and the vulnerability hidden in the contract may not be detected.

\subsection{Language Translation}

Because there are no related models or tools designed for smart contract platform, both symbolic execution and formal verification require language translation to transform the unfamiliar languages into a familiar one.  

For example, in symbol execution, its core model `SMT solver' has many existing tools (such as Z3, SMT-ART, etc.) available, but these tools only support specific languages like C++, Java or LLVM bytecode. This means that if we want to use these tools, we must first convert the contract code written by solidity into the language corresponding to the tools. However, due to some unique features in the solidity language, converting it to another language usually requires additional manipulation of some of the instructions. For example, the invocations in solidity can be divided into three types: internal, external, and \textbf{call()}. Internal calls and external calls exist in most programming languages, but the \textbf{call()}, which can call methods in other contracts, is rare in other programming languages and requires special handling when translating. 
Formal verification also faces the same problem. Often, when we use model checking to guide the development of an application, we need to convert the requirements into specifications using abstract language or paradigm. After passing the detection of the model detector, we can convert the specifications into application code. When we need to verify that an application has a problem, we need to convert the code to specifications first, but at this time, the conversion may encounter many problems. Bhargavan`s team \cite{bhargavan2016formal} wants to convert the code in solidity to F* and also decompile bytecode in EVM to F* then verify their equivalence. However, in the evaluation part, they also mentioned that F* does not support many of the syntactic features in Solidity, resulting in only 46 of the 396 contracts to be translated. 

In general, both the symbolic execution and the formal verification are developed well in software vulnerability mining technology. For the new platform of smart contracts, the challenge in applying this technology is how to translate Solidity into the required language fully.

\subsection{Analysis}

Whether it's fuzzing, symbolic execution or formal verification, they end up using static analysis methods when mining vulnerabilities. In \cite{jiang2018contractfuzzer}, for the 7 different vulnerabilities proposed, they design corresponding oracles to determine whether the contract is vulnerable when analyzing the fuzzing execution results. For example, for the gasless problem, if the gas value of the \textbf{call()} function in the contract is 0 during the analysis, the contract is considered to have a gasless problem. Here, whether the remaining number of gas of the \textbf{call()} in the contract is 0 will be considered as a constraint for detecting this vulnerability. Such constraints make the analysis simple and easy to automate, but also bring false positives and false negatives. 

The highly customized static analysis method also brings limitations to scalability. The existing analysis aims at some vulnerabilities that have been studied, so these methods cannot identify other vulnerabilities that are not discovered or appear in unnoticed scenarios. Of course, a good constraint can help us cover multiple manifestations of the same vulnerability, but whether we recognize all the patterns of this vulnerability depends on the triggering logic of the vulnerability itself. 

In a nutshell, how to design a constraint that captures most of the vulnerability patterns is a challenge for these static analysis methods.

\section{Conclusion}

In this paper, we survey several vulnerabilities focused on smart contracts in Ethereum such as reentrancy attacks, gasless send and vulnerability about timestamps. Then we present their triggering logic. Also, we briefly introduce fuzzing, symbolic execution, formal verification, and language translation methods that are often used for software vulnerability detection, and overview some tools for applying these methods to smart contract vulnerability detection.

These software vulnerability detection tools excel in detecting vulnerabilities in smart contracts, but at the same time, for such a new platform, the new features in smart contracts also pose challenges to the application of these tools. For different tools, we analyze the limitations imposed by the new features of Ethereum and raise the challenges that may constrain their development.



\bibliographystyle{IEEEtran}
\bibliography{bib}

\end{document}